\documentclass[letter]{aa} % for the letters 
\usepackage{graphicx}
\usepackage{txfonts}
\begin{document}
   \title{Exoplanets transmission spectroscopy: accounting for eccentricity and longitude of periastron}

   \subtitle{Superwinds in the upper atmosphere of HD209458b?}

   \author{M. Montalto
          \inst{1},
	   N. C. Santos
          \inst{1,2},
	   I. Boisse
          \inst{1},
           G. Bou\'e
          \inst{1},
	   P. Figueira
          \inst{1},
           S. Sousa
           \inst{1}
          }

   \institute{Centro de Astrofisica da Universidade do Porto,
              Rua das Estrelas 4150-762, Porto, Portugal.\\
              \email{Marco.Montalto@astro.up.pt}  
            \and  
              Departamento de Física e Astronomia, Faculdade de Ciências,  Universidade do Porto,
              Portugal\\
             }

   \date{}

% \abstract{}{}{}{}{} 
% 5 {} token are mandatory
 
  \abstract
   {
    A planet transiting in front of the disk of its parent star offers the
    opportunity to study the compositional properties of its atmosphere
    by means of the analysis of the stellar light filtered by the planetary
    atmospheric layers. Several studies have so far placed useful
    constraints on planetary atmospheric properties using this technique,
    and in the case of HD209458b even the radial velocity of the planet during
    the transit event has been reconstructed opening a new range of possibilities.
   }
   {
    In this contribution we highlight the importance to account for the orbital 
    eccentricity and longitude of periastron of the planetary orbit to accurately 
    interpret the measured planetary radial velocity during the transit.
   }
   {
    We calculate the radial velocity of a transiting planet in an eccentric orbit. 
   }
   {
    Given the larger orbital speed of planets with respect to their stellar companions 
    even small eccentricities can result in detectable blue or redshift radial velocity offsets 
    during the transit with respect to the systemic velocity, the exact value depending also on the
    longitude of the periastron of the planetary orbit. 
    For an hot-jupiter planet, an eccentricity of only $e$=0.01 can produce a radial velocity 
    offset of the order of the km/s. 
   }
   {
     We propose an alternative interpretation of  
     the recently claimed radial velocity blueshift 
     ($\sim2$ km/s) of the planetary spectral lines of HD209458b which implies
     that the orbit of this system is not exactly circular. In this case,
     the longitude of the periastron of the stellar orbit is
     most likely confined in the first quadrant (and that one of the planet in the 
     third quadrant). We highlight that transmission spectroscopy allows not only to 
     study the compositional properties of planetary atmospheres, 
     but also to refine their orbital parameters
     and that any conclusion regarding the presence of windflows on planetary surfaces 
     coming from transmission spectroscopy
     measurements requires precise known orbital parameters from RV.
   }


   \maketitle
%
%_______________________________________________________________________

\section{Introduction}

The first proof of the existence of an exoplanetary atmosphere was 
given by Charbonneau et al.~(2002), 
once they succeeded in the detection of sodium absorption 
in the transmission spectrum of the exoplanet
HD209458b. This result was obtained using the 
$Space$ $Telescope$ $Imaging$ $Spectrograph$
($STIS$) on board of the $Hubble$ $Space$ $Telescope$ ($HST$).
In the following years several other attempts were made
to detect absorption features from exoplanets atmospheres 
using also ground-based observatories
but, until recently, they were able to place only upper limits
(Moutou et al.~2001; Snellen~2004; Narita et al.~2005).
Sodium detection from the ground was first achieved for HD189733b by 
Redfield et al.~(2008) and later confirmed in HD209458b by 
Snellen et al.~(2008) who found Na levels which matched $HST$ 
values found by Sing et al.~(2008). Additionally Sing et al.~(2010)
using ground-based narrowband spectrophotometric measurements at the $GTC$,
detected potassium absorption in XO-2b,
while Wood et al.~(2010) detected sodium absorption in WASP-17b 
by means of transmission spectroscopy at the $VLT$. 
Recently Snellen et al.~(2010) presented further ground-based 
detection of $CO$ absorption lines in the atmosphere of HD209458b.
Their refined analysis allowed to isolate for the first time 
the doppler shift of the planetary spectral lines during the 
transit, allowing a direct determination of the masses both of the
star and the planet, in the same manner as done for double lined 
eclipsing binaries. In the course of their analysis, 
Snellen et al.~(2010) also noticed that the $CO$ planetary absoption
lines appeared blueshifted with respect to the systemic velocity of 
the host star, a fact that was attributed to the presence of superwinds
on the planetary surface, flowing from the day to the dark side of
the planet and crossing both its equatorial and polar regions.
However, the planetary orbit of HD209458b was assumed to be perfectly 
circular. 
The aim of this contribution is to analyze which consequences has the 
presence of a non-null
planetary orbital eccentricity on transmission spectroscopy measurements.
Even if a residual small orbital eccentricity is present the velocity of the
planet during the transit can be expected to be offset with respect
to the systemic velocity, and by a significant amount given
the large orbital speed of the planet, as demonstrated in 
Sect.~\ref{sec:model}. 
In Sect.~\ref{sec:HD209458b}, we discuss the particular case of HD209458b.
Finally in Sect.~\ref{sec:conclusion}, we summarize our results and
conclude.

%_______________________________________________________________________

\section{Transmission spectroscopy of an eccentric transiting planet}
\label{sec:model}

We assume a two body system composed of a planet and its host star.
The radial velocity $RV$ of the planet with respect to the barycenter of the
system is given by (e.g. Hilditch 2001):

\begin{equation}
RV=\frac{(2\pi)^{1/3}}{P^{1/3}}\,\frac{(G)^{1/3}m_s\,sin(i)}{\sqrt{1-e^2}\,(m_s+m_p)^{2/3}}\,\Big(cos(\omega+f)+e\,cos\omega\Big),
\end{equation}

\noindent
where $f$ is the true anomaly, $e$ is the eccentricity of the orbit, and $i$ is the inclination with respect to
the plane of the sky, $m_s$ and $m_p$ are the masses of the star and of the planet,
$\omega$ is the longitude of the periastron of the planetary orbit, $P$ is the orbital period, and $G$ the gravitational constant. 
Isolating the terms dependent on the eccentricity and the longitude of the periastron
and grouping all the others in the constant $\tilde K$ we obtain:

\begin{equation}
RV={\tilde K}\,\frac{cos(\omega+f)+e\,cos\omega}{\sqrt{1-e^2}}.
\end{equation}

\noindent
Once the planet crosses the line of sight of the observer the term dependent
on the true anomaly is exactly null, then the radial velocity of the
planet in that moment ($RV_0$), with respect to the barycentric radial
velocity, is given by:

\begin{equation}
RV_0={\tilde K}\,\frac{e\,cos\omega}{\sqrt{1-e^2}}.
\end{equation}

\noindent
This velocity is null only if the eccentricity is null or if the longitude
of the periastron equals 90$^{\circ}$ (or 270$^{\circ}$). 
Assuming $m_s=1\,M_{\odot}$,
$m_p=1\,M_{jup}$, $i=90^{\circ}$, $P=3$ days and $e=0.01$ we can derive the following
upper limit for $RV_0$ given that $|cos\omega|\le$1:

\begin{equation}
|RV_0|\,\leq\,1.48\,\rm km/s.
\end{equation}

\begin{figure}
\center
\includegraphics[width=\columnwidth]{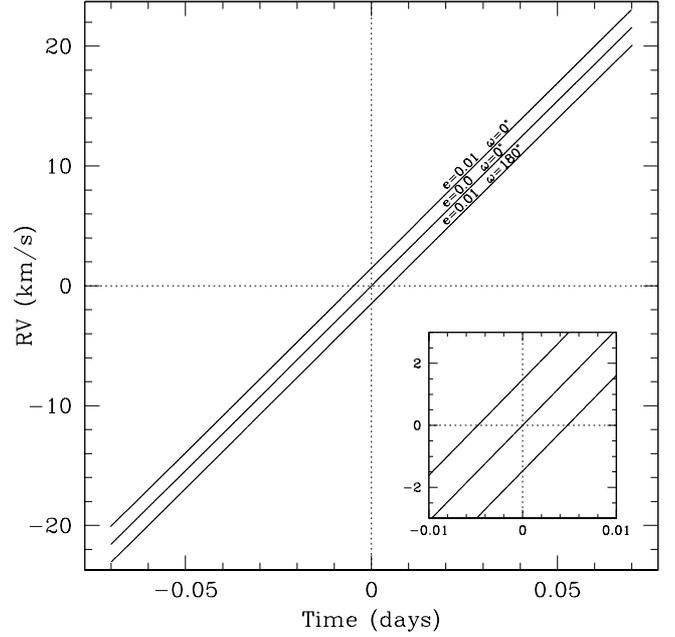}
    \caption{
             The continuous lines represent the planetary radial velocity 
             during the transit for different
             values of the eccentricity ($e$) and the longitude of periastron
             ($\omega$). 
             In this figure we assumed $m_s=1\,M_{\odot}$, $m_p=1\,M_{jup}$, $i=90^{\circ}$, 
             $P=3$ days. The dotted lines indicate the mid-transit point and the systemic velocity.
             The box in the lower right corner is a magnified view close to the mid-transit point.
            }
\label{fig:off}
\end{figure}

\noindent
This result is totally due to the fact that the orbital speed of the planet is
much larger than that one of its stellar companion, given its much
smaller mass. Since transmission spectroscopy allows to sample only a small portion of the
planetary orbit close to the mid-transit point, all the measurements acquired
during the transit would appear to be offset with respect to the systemic
velocity as represented in Fig.~\ref{fig:off}.
Then, considering the case of an hot-jupiter, even a small eccentricity can produce
a radial velocity offset of the order of the km/s (depending also on the value of the
longitude of the periastron). This offset can be either blueshifted or
redshifted. If the eccentricity of the system is completely neglected, assuming
that the orbit is exactly circular, this radial velocity offset
may be interpreted as having a different physical origin.

%__________________________________________________________________

\section{The case of HD209458b}
\label{sec:HD209458b}

As reported in the introduction, transmission spectroscopy 
was applied to the case of HD209458b by several authors in the
past placing useful constraints on the abundances of different
elements in its atmosphere. Recently Snellen et al.~(2010),
presented a refined procedure by means of which the orbital motion
of the planet during the transit was unveiled for the first time 
throught the analysis of the doppler effect of $CO$ absorption lines.
The signal they detected appeared blueshifted with respect to the
systemic velocity of the host star by around 2 km/s and the uncertainty
estimated by the authors was 1 km/s. 
The authors then interpreted this blueshifted signal
as the evidence that superwinds are flowing on the surface
of the planet, from the dayside to the darkside and 
crossing both the equator and the poles.

This conclusion was drawn on the basis of the assumption that
the orbit of HD209458b is perfectly circular. High precision
radial velocities allow to conclude that the eccentricity of this system
is consistent with zero within the uncertainties (Laughlin et al.~2005).
In particular, Laughlin et al.~(2005) pointed out that:
{\it ``even when the orbit underpinning a data set is circular, e computed from 
an ensemble of bootstrap trials will have a characteristic nonzero value''.}
Nevertheless considering the uncertainties a residual small eccentricity cannot be ruled out a priori
(Kipping 2008) and in the last years this was in fact a matter of large debate.
Winn et al.~(2005), analyzing high-precision
radial velocities of the host-star, photometry and timing of the secondary eclipse,
obtained for the eccentricity and longitude of periastron of the star ($\omega_s$) the values
$e=0.0147\pm0.0053$ and $\omega_s=84^{\circ}\pm11^{\circ}$. If these mean values
are used in Eq.~3, we obtain an expected blueshifted signal equal to 
$RV_0=0.21$ km/s. However, considering the 90$\%$ upper confidence limits ($e\,cos\omega=0.0049$)
reported by Winn et al.~(2005) we derive that $RV_0$ could be as high as
 $RV_0=0.68/\sqrt{1-e^2}$ km/s. Similarly, 
 Deming et al.~(2005) derived that the secondary eclipse occurs at the midpoint
between transits within 21 min (3-$\sigma$). This
translates into an upper limit for the expected radial velocity offset 
equal to $RV_0=0.91/\sqrt{1-e^2}$ km/s.
These estimates appear still consistent with the observed blushifted signal
considering the uncertainties of the observations.

Then at present, given the large sensitivity
of the mechanism here presented on the orbital eccentricity, and given the uncertainty
of the measurements, it is questionable if other mechanisms like superwinds 
should be invoked to account for the observations. As explained by Deming et al.~(2005),
 even a dynamically significant eccentricity ($e\sim0.03$) could be
still in agreement with their 3-$\sigma$ limit,
despite requiring a rather good alignment of the apsidal line with the line of sight
$|\omega-\pi/2|<12^{\circ}$.

\begin{figure}
\center
\includegraphics[width=\columnwidth]{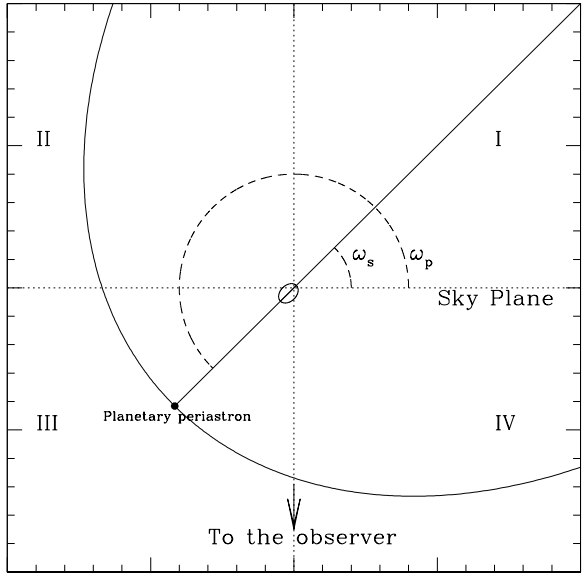}
    \caption{
             Schematic representation of the orbit of the planet (large ellipse)
             and the orbit of the star (small ellipse), with the definition of
             the quadrants and of the longitude of periastron of the star ($\omega_s$)
             and of the planet ($\omega_p$). The dotted horizontal line indicates
             the plane of the sky, whereas the dotted vertical line the direction to the observer
             (arrow).
            }
\label{fig:fig2_orbits}
\end{figure}

If totally attributed to the eccentricity, the observed blueshift would
imply that the longitude of the periastron of the planet ($\omega_p$) should
lie either in the second or in the third quadrant, and consequently that one 
of the star ($\omega_s=\omega_p+180^{\circ}$) either in the fourth or in the 
first quadrant respectively (see Fig.~\ref{fig:fig2_orbits}). 
Despite the large uncertainty,
Winn et al.~(2005) report $\omega_s=84^{\circ}\pm11^{\circ}$ (1-$\sigma$).
Taken together, the above considerations reinforce the idea 
that the longitude of the periastron of the star should lie most likely 
in the first quadrant (and that one of the planet in the third).

%__________________________________________________________________

\section{Conclusions}
\label{sec:conclusion}

Thanks to the large orbital speed of planets with respect to their stellar companions,
transmission spectroscopy allows not only to constrain the properties
of the atmosphere of a transiting planet, but also offers an alternative mean
to refine its orbital parameters. 
For the eccentric planet Gj436b ($P=2.6438986$ days, $e=$0.15, $\omega_s=351^{\circ}$)
we expect a redshifted radial velocity offset equal to 11.68 km/s. 
For the particular case of HD209458b, nominally almost half of the blueshifted
signal reported by Snellen et al.~(2010) could be explained
assuming that the orbital eccentricity of the system
is not exactly null (just at the percent level) considering the limits 
imposed by transit timing of the secondary eclipse given by Deming et al.~(2005).
However, given the uncertainty of the measurements of Snellen et al.~(2010),
our estimates appear still consistent with the observations.
Once attributed to the eccentricity, the blueshifted 
signal together with radial velocity measurements of the host star,
allow to confine the longitude of the periastron of the star
in the first quadrant (and that one of the planet in the third quadrant).

Finally, it should be also noted that any conclusion regarding the
presence of windflows on planetary surfaces coming from transmission spectroscopy
measurements requires precise known orbital parameters from RV.

%__________________________________________________________________

\begin{acknowledgements}
This work was supported by the European Research Council/European Community under the FP7
through Starting Grant agreement number 239953, and throuth grants reference
PTDC/CTE-AST/098528/2008 and PTDC/CTE-AST/098604/2008 from Funda\c{c}\~ao para a
Ci\^encia e a Tecnologia (FCT, Portugal). MM, NCS, PF and SS also acknowledge the support from FCT
through program Ci\^encia\,2007 funded by FCT/MCTES (Portugal) and POPH/FSE (EC) and in
the form of grants reference PTDC/CTE-AST/66643/2006, PTDC/CTE-AST/098528/2008, and
SFRH/BDP/71230/2010.
\end{acknowledgements}

%__________________________________________________________________

%__________________________________________________________________

\end{document}